# An Algorithmic Structuration of a Type System for an Orthogonal Object/Relational Model

Amel Benabbou
Department of Computer science
Es-Sénia University
Oran 31000 Algeria
benabbou_amel@yahoo.fr

Safia Nait Bahloul
Department of computer Science
Es-Sénia University
Oran 31000 Algeria
nait1@yahoo.fr

Youssef Amghar
LIRIS, Batiment Blaise Pascal
7 Avenue Jean Capelle
69621 Villeurbanne Cedex
youssef.amghar@insa-lyon.fr

**Abstract**

Date and Darwen have proposed a theory of types, the latter forms the basis of a detailed presentation of a panoply of simple and complex types. However, this proposal has not been structured in a formal system. Specifically, Date and Darwen haven't indicated the formalism of the type system that corresponds to the type theory established. In this paper, we propose a pseudo-algorithmic and grammatical description of a system of types for Date and Darwen's model. Our type system is supposed take into account null values; for such intention, we introduce a particular type noted #, which expresses one or more occurrences of incomplete information in a database. Our algebraic grammar describes in detail the complete specification of an inheritance model and the subryping relation induced, thus the different definitions of related concepts.

## I. Introduction

A type theory is the basis of a formal type system that can serves as an alternative to the naive theory of sets [1, 2], it allows an expressive specification for languages and constitutes a powerful tool in logic demonstration [3]. This could be described precisely in the framework of a $\lambda$ cube [4, 7] or a pure type system [5]; for a general framework, we refer to [6, 8]. From a practical point of view, a theory of types provides a formal basis for the study, conception, and analysis of type systems [9].

Originally, a minimal type system was proposed by Church in [4], for witch mathematical objects are of two sorts: terms and types, the terms of the type system are $\lambda$-calculus terms, which have become a formalization of the partial recursive functions proposed in [10]. This type system has been generalized for other simple data types such cartesian product and disjoint union [9], an extension of this system was accomplished by Girard in [11].

A type system is a form of Meta information that provides a typical structure of a language or a model; it is represented by a formal syntax to concisely define the set of rules allowing to infer a collection of simple or complex types or to specify concepts, expressions and queries of a given language. These typing rules define how the expressions of a language are classified in terms of type and values [12] allowing to avoid inconsistency and shall maintain a correct model for the whole. The rigorous definition of the type system is a crucial step for determining a coherent model and, by allowing in particular its abstraction and semantic specification. However, many works were intended for typing and corresponding problems [3, 6, 7, 13, 14, 15, 16].

Currently, new approaches are proposed to include other concepts in the context of programming languages with dependent types and databases models, such have been defined in [5.17, 18]. Nevertheless, most studies have focused on defining a type system for object-oriented language [8, 19, 20], for the script languages [12, 21, 22] and type systems for generic programming languages by constraints [23, 24, 25]. More recently, works present type systems for the semantic Web languages [26, 27, 28] designed using typing rules inspired from [9], it is a reformulation of the type system presented in previous works [29.30].

In [31.32], Date and Darwen have proposed a theory of type which is the basis of a panoply of simple and complex types. Although, this definition has not been structured in a formal system. Specifically, Date and Darwen have not indicated the formalism of the type system that corresponds to the theory of type established.

Indeed, this paper is a formal description of the type system for the model of Date and Darwen, taking into account their inheritance model. To do this, the article is organized as following: The next section is an overview of the model of Date and Darwen, including definitions of key concepts of the theory of type added. An exhibition of the inheritance model is made in Section 3. Section 4 is intended for the definition and treatment of the formal model of Date and Darwen. The end of the paper is signed by a conclusion.

## II. Orthogonal model of Date and Darwen

Date and Darwen [31.32] propose a theoretical basis for integration of a few objects concepts in the relational context; they consider that the essential concepts of object orientation are orthogonal with respect to the theoretical foundations of the relational model of Codd [33.34]. Thus, it is not necessary to impose an extension to this model to achieve the object concepts. It is enough expand domains to new simple or complex user types, and allow inheritance and subtyping in order to take advantage of the characteristics of object-orientation (eg polymorphism) and facilitate reuse.

The latest edition of the third manifesto of Date and Darwen [31] stand out the two key features: a strong focus on the relational model and a thorough treatment of type theory.

## II.1. Date and Darwen's Vision of relational model

According to Date and Darwen, the model proposed in [31.32] is a meticulous and accurate description of the "classical" version of relational model [33.34]. It includes the basics of relational theory like: *attribute, relation, tuple*, etc.... In that data model, Date and Darwen have given formal definitions adapted to their vision of future databases that incorporates to the relational model, concepts issue from the object context. The next section is devoted to such definitions:

**Definition 1 (Heading).** A *Heading* {H} is a set of ordered pairs <A, T> such as:
  a. A is the name of attribute in {H}.
  b. T is the declared type of the attribute A.
  c. Two pairs $<A_1, T_1>$ and $<A_2, T_2>$ are such that $A_1 \neq A_2$
A Heading {H} of degree zero is valid, and it called the empty heading { }.

*Example*
  {< S #, S#>, < SNAME, NAME >, < STATUS, INTEGER >, < CITY, CHAR >}
      /* A heading $\{H_1\}$ of a relation R*/

**Definition 2 (Tuple).** Given a collection of types $T_i$ (i = 1, 2, ..., n, where n ≥ 0), not necessarily all distinct, a *tuple* t -over those types- is a set of n ordered triplets of the form $<A_i, T_i, v_i>$, such as vi is the value of the attribute $A_i$ of type $T_i$.
A tuple of degree zero is also valid, and it's called the empty tuple noted 0-tuple = TUPLE { }.

*Example*
{< S #, S#, S1>, < SNAME, NAME, SMITH >, < STATUS, INTEGER, 20 >, < CITY, CHAR, LONDON >}
      / * A tuple that conform to the heading $\{H_1\}$* /

**Definition 3 (Body of relation).** A *body* Br of a relation r is a set of tuple $t_i$. However, there may be exist tuples tj that conform to the heading {H} without that tj $\in$ B.

*Example*
  $t_1$: Tuple {S # S # ('S1'), SNAME NAME ('Smith'), STATUS 20, CITY 'London'},
  $t_2$: Tuple {S # S # ('S2'), SNAME NAME ('Jones'), STATUS 10, CITY ' Paris'}.

      / * $t_1$ is a tuple in the relation R, $t_2$ can be a tuple in the relations S and R that
              both conform to the heading $\{H_1\}$* /

**Definition 4 (Relation).** A *relation r* is defined by its heading $\{H_r\}$ and its body $B_r$. The Heading {H} represents the schema of the relation *r*.
A relation of degree zero is valid, and there exist two such relations:

  a. TABLE_DEE that contains only the tuple 0-tuple : r = RELATION { } {Tuple { }}
  b. TABLE_DUM that contains no tuple : r = RELATION { } { }.

## II.2. Type theory of Date and Darwen model

Support for some features of object-orientation [19], brings many advantages to relational model [33.34], these features are orthogonal to the relational model. Otherwise, the relational model does not need any revision or correction, rather extend the types to: user types (abstract data types) [2, 8, 13], and then allow subtyping [35, 36, 37, 38] between such types [39, 40]. On the other hand, this may be the basis of the definition of a database language that: (a) is relational, (b) take account of these characteristics orthogonal, and (c) may represent the basis foundations of the model. Such language is known as the *language D* (e.g. any language that conforms to the principles of third manifesto). In this optic, Date and Darwen have defined such language, and have called Tutorial D. Thereafter, all types definitions are described in
*Tutorial D*.

Therefore, we present the main types introduced into the type theory associated to the data model of Date and Darwen, with their proposed inheritance model.

### II.2.1 Scalar type

It seems to be very difficult to come up with a definition of the term *scalar* that is both precise and useful; indeed, [41] shows that the concept of atomicity – which is just the concept of 'scalarness' by another name – has no absolute meaning. For example, consider the string 'New York' (a value of type CHAR), if this string is considered to be a single value, then it seems reasonable to think of it as scalar; but if it is considered to be composite (consisting as it does of sequence of either words or letters), then it seems reasonable to think of it as nonscalar.

**Definition 5 (Scalar type).** A type is said scalar if it has no *user- visible components* (attributes). Given two distinct scalar types $T_1$ and $T_2$, with the corresponding sets of values $S_1$ and $S_2$, respectively; then:

   a. The names of types $T_1$ and $T_2$ are distinct
   b. $S_1$ and $S_2$ are disjoint.

The values of each type are characterized by a exactly one *physical representation* (encoding internal: decimal, binary ...) and one or more *possible representations*. To be more specific, Date and Darwen have used the term scalar type to mean a type that is neither a tuple type nor a relation type, and the term nonscalar type to mean a type that is a tuple or relation type.

A scalar type can be defined by user (*User-defined scalar type*), or provided by the system (*Built-in* or *System-defined scalar types*):

**a) User-defined scalar type**

Let T a scalar type, and v a value of type T. By definition, the *physical representation* of these values of type T is hidden from users; however, all declared possible representations (abbreviated POSSREP) have components that are visible to users. For example, the user-defined type POINT is defined as follows:

```
TYPE POINT /* geometric points in two- dimensional space*/
         POSSREP  CARTESIAN  {X RATIONAL, Y RATIONAL}
         POSSREP POLAR {R RATIONAL, THETA RATIONAL}.
```

The type POINT has two Possreps CARTESIAN and POLAR, which describe the fact that a point can be represented by Cartesian or polar coordinates. Each Possrep has two components X and Y of rational types witch are visible to users.

**b) Built-in Scalar type (or system- defined Scalar type)**

It's the set of types constructed by the system, such a type can be: INTEGER, RATIONAL, CHARACTER and BOOLEAN.

Generally, a system- defined type has no Possrep. However, some types can have a possible representation just as user- defined types do. Type DATE (Gregorian dates) with the Possrep {YEAR INTEGER, MONTH INTEGER, DAY INTEGER …} might be an example; type Complex (complex numbers) might be another. For reasons of simplicity, we consider in our article, that all system- defined types have no Possreps.

### II.2.2. Nonscalar types

According to Date and Darwen [31, 32], the nonscalar term is used to designate the tuple / relation types. Thus, we discuss the types which are obtained by invoking the TYPE GENERATOR (or type constructor) TUPLE and RELATION. Remember that all the *components* of a nonscalar type are not visible.

**a) Type Generator Tuple**

Given some heading {H}, the *generated type* TUPLE {H} is the *tuple type*. That type is considered as a basis for defining variables and operators of type tuple. In *Tutorial D*, tuple types can simply be used typically as part of the operation that defines a tuple variable of type TUPLE {H} which is the name of the tuple type in question. In fact, *Tutorial D* deliberately does not provide any kind of explicit" define tuple type" operator.  Indeed, it is possible to define a tuple variable as follow:

```
   VAR ADDR TUPLE {S# S#, SNAME NAME, STATUS INTEGER, CITY CHAR}
/* ADDR is a variable of type TUPLE {STREET CHAR, CITY CHAR,
STATE CHAR, ZIP CHAR} */
```

The Heading {H} is recursively defined, for example, the following statement defines a variable tuple, and one of its attributes is of type tuple:

```
 VAR NDDR2 TUPLE {NAME NAME, ADDR TUPLE {STREET CHAR, CITY CHAR,
                   STATE CHAR, ZIP CHAR}});
 /* ADDR is an attribute of type TUPLE {STREET CHAR, CITY CHAR,
                   STATE CHAR, ZIP CHAR}*/
```

**b) Type Generator Relation**

Given some heading {H}, the *generated type* RELATION {H} is the relation type. Any language D must support the use of the generated type RELATION {H} as a basis to define variables and operators of that type. In *Tutorial D*, relation types can simply be used typically as part of the operation that defines a relation variable "relvar" of type RELATION {H} which is the name of the relation type in question. Such variable is defined as:

```
VAR S RELATION {S # S#, SNAME NAME, STATUS INTEGER, CITY CHAR};
 /* S is a relvar of type RELATION {S # S#, SNAME NAME, STATUS
                     INTEGER, CITY CHAR}*/
```

An attribute can be of scalar or nonscalar type. Thus a relation may have attributes of witch values are of tuple or relation types, for example:

```
    VAR SPQ…RELATION {S# S#, PQ RELATION {P# P#, QTY QTY}};
/* SPQ is of type RELATION {S# S#, PQ RELATION {P# P#, QTY QTY}*/
   /* PQ is an attribute of type RELATION {P# P#, QTY QTY}}*/
```

According to Date and Darwen, the concept of nested headings which can be defined recursively in terms of themselves still an open question. However, this notion is not allowed in their model.

The type theory proposed by Date and Darwen in [31] is the basis of an inheritance model and subtyping relation between the types. Indeed, we address in what follows, the inheritance model proposed by the both authors.

### III. Date and Darwen Inheritance model

According to classical notion, the term *inheritance* refers to the phenomenon by which we can say, for example, that each circle is an ellipse, from which all the properties applied to the ellipses in general are also applicable to circles.

The relation of subtyping is the basic concept for conception of an inheritance model [31.39]. Particularly, this concept was introduced by Cardelli in [35] and Taivalsari [38], allowing to construct inheritance systems between types, which are more expressive in order to express the inclusions in sets and sub sets [25].

The inheritance model of Date and Darwen [31, 32] distinguishes two categories of inheritance: single [31, 37] and multiple [31, 35, 36]. Precisely, single inheritance is a special case of multiple inheritances. In this regard, Date and Darwen have considered the following approach:

1) First, construct a sound model of single inheritance.
2) Extend that model to incorporate multiple inheritances subsequently.

Commonly, the term' inheritance' is applied on values of *scalar* type; whatever, in the model of Date and Darwen, it also implies the *nonscalar* type, i.e. tuple and relation types. Thus, we present the new concepts and terms introduced in such inheritance model:

1. Every type is a *supertype* itself (e.g., type ELLIPSE is a supertype of ELLIPSE).

2. If T is a supertype of T ', T and T' are distinct, then T is a *proper supertype* of T' (e.g., POLYGON is a proper supertype that type of SQUARE).

3. Every type is a *subtype* of itself (e.g., ELLIPSE is a subtype for ELLIPSE).

4. If T' is a sub type of T, T' and T are distinct, then T' is a *proper subtype* of T (e.g., SQUARE is a proper subtype of POLYGON).

5. If T is a *proper supertype* of T', and there is no type that is both a *proper subtype* of T and a *proper supertype* of T', then T is an *immediate supertype* of T' and T' is an *immediate subtype* of T (e.g., RECTANGLE is an *immediate supertype* of SQUARE, and SQUARE is an *immediate subtype* of RECTANGLE).

6. A *root type* is a type with no proper supertype (e.g., FIGURE is a root type); a *leaf type* is a type with no proper subtype (a.g., SQUARE is a leaf type). *Note: strictly speaking, a given type can be said to be root or leaf type only in the context of some specific type hierarchy or type graph. For example, type RECTANGLE is a leaf type in the hierarchy that results from by deleting type SQUARE from the corresponding hierarchy.*

7. If T is a supertype of T ', then every value v has exactly one *most specific type*. In this case, a given value v can be of type T, and not of type T', this means that T is its *most specific type*, noted (MST (v)). For example, a value $v_1$ might be ' just an ellipse' and not a circle, meaning its MST(v) = ELLIPSE.

**III.1. Single inheritance**

According to Date and Darwen specification, a *single inheritance* is an hierarchy in which every *proper subtype* has exactly one *immediate supertype*. We address the following section to discuss single inheritance for *scalar* and *nonscalar* types.

**III.1.1. Scalar type**

A single inheritance [31, 35] on scalar type is represented by a directed acyclic type graph. For example, the type RECTANGLE is a subtype of the super POLYGON; all polygons properties are inherited by rectangles. In Tutorial D, and CIRCLE and ELLIPSE types are defined as follows :

```
    TYPE ELLIPSE
        IS {FIGURE
            POSSREP {A LENTH, B LENTH, CTR POINT
                     CONSTRAINT A≥B}}
```
/*ELLIPSE is a subtype of FIGURE, the possrep is described pa the constraint a≥b*/

```
    TYPE CIRCLE
        IS {ELLIPSE
            CONSTRAINT THE_A (ELLIPSE) = THE_B (ELLIPSE)
            POSSREP {R = THE_A (ELLIPSE),
                     CTR = THE_CTR (ELLIPSE)}};
```
/* CIRCLE is a subtype of ELLIPSE, with additional constrain A=B, its possrep is derived from that of type ELLIPSE*/

A type in an inheritance context might be a *union* or *nonunion* type, that concept is necessary for the definition of our type system, therefore we retain the following definitions:

**Definition 6 (union type).** Let T a scalar type and V the set of values v of type T, type T is said union (abstract type) if and only if all possible values v of type T are also of an immediate subtype of T.

This means, that there is no value $v \in V$ such that MST (v) = T. Thus, a *union* type requires the existence of at least two *immediate subtypes*. The utility of such type is specifically to avoid redundancy during the definition of operators that applied on proper subtypes for the union type in question, for example:

```
TYPE NONCERCLE
    IS {ELLIPSE
        CONSTRAINT THE_A (ELLIPSE) >THE_B (ELLIPSE)
        POSSREP {A = THE_A (ELLIPSE),
                 B = THE_B (ELLIPSE),
                 CTR = THE_CTR (ELLIPSE)}};
```
/* NONCERCLE is an immediate subtype of the type ELLIPSE*/

Then, the type ELLIPSE is the *union* type of both types CIRCLE and NONCIRCLE, a new definition type ELLIPSE can be:

```
TYPE ELLIPSE UNION
     IS {FIGURE
         POSSREP {A LENTH, B LENTH, CTR POINT
                  CONSTRAINT A≥B}};
```
/* the UNION specification is added to the new definition of type ELLIPSE*/

Furthermore, the *union* type involves the definition of a special for witch the possible representation is missing, this concept is the subject of the following definition.

**Definition 7 (Dummy type).** A dummy type is a union type that has no declared possible representation possrep; a given union type shall be permitted to be a dummy type if and only if it is empty or it has no regular immediate supertype (the alpha and omega types, respectively, as defined later), for example:

```
TYPE ELLIPSE UNION
         IS {FIGURE};
```
/* ELLIPSE is now a dummy, because it has no declared possrep nor contraint*/

Conceptually, Date and Darwen have introduced two special dummy types:

1. *Type Alpha* : is the *maximal* type with respect to every scalar type, that contains all scalar values and is a supertype of every scalar type; more precisely, it is a proper supertype of every scalar type except itself, and an *immediate supertype* of every scalar type that would otherwise be a *root* type; by definition, *alpha* has no declared possible representation and no *immediate supertype*.

2. *Type Omega* : is the *minimal* type with respect to every scalar type, it contains no values at all and is a subtype of every scalar type; more precisely, it is a proper subtype of every scalar type except itself, and an *immediate subtype* of every scalar type that would otherwise be a *leaf* type; by definition, *omega* has no declared possible representation and no *immediate subtype*.

**III.1.2. Nonscalar type**

The definition of inheritance relation between nonscalar types is recursive insofar as tuple and relation tupes are based on the concept of subtyping between *scalar* types. Note that all the specifications of the two nonscalar types in this section are given in the same definitions. Thus, we say that the tuple / Relation type T', with the heading:

$$\{<A_1, T'_1>, <A_2, T'_2>,..., <A_n, T'_n>\}$$

Is a *subtype* of the type T (or T is a supertype of T '), with the heading:

$$\{<A_1, T_1>, <A_2, T_2>,..., <A_n, T_n>\}$$

If and only if, for all i (i = 1,2, ..., n), the type $T_i'$ is a subtype of the type $T_i$ (or $T_i$ is a supertype of $T_i'$). For example, we consider the following relation types:

      RELATION {E ELLIPSE, R RECTANGLE}      /* Relation type "ER" */
      RELATION {E CIRCLE, R SQUARE}          /* Relation type "CS" */

Type CS is an *immediate subtype* of the type ER, since types CIRCLE and SQUARE are *immediate subtypes* of types ELLIPSE and RECTANGLE, respectively.

**a. Union and dummy types**

The notion of union type on nonscalar types are indeed related on that on attributes types witch constitute the heading {H}of the corresponding tuple/relation type in order to verify the definition (6) and (7). Therefore, we say that the tuple /relation T, with the heading:

$$\{<A_1, T_1>, <A_2, T_2>,..., <A_n, T_n>\}$$

Is said Union, if and only if, for all i (i = 1,2, ..., n), the type $T_i$ is a union type. For example, if ELLIPSE is a union type, then the tuple type TUPLE {E ELLIPSE, X CHAR} and relation type RELATION {E ELLIPSE, X CHAR} are also union types.

In the same way, this definition may satisfy the definition of dummy type for both nonscalar tuple and relation types, in which every type in the heading {H} is a dummy type. For example, if ELLIPSE is a dummy type, then so is TUPLE {E ELLIPSE, X CHAR} and so is RELATION {E ELLIPSE, X CHAR}.

**b. Minimal and maximal types**

As for scalar types, the definition of dummy type involves two particular types :

1. *T/R_alpha type*: the type T/R_alpha is said maximal type with respect to a *tuple /relation* type T, with headings, respectively:

$$\{<A_1, T_1\_alpha >, <A_2, T_2\_alpha>,.., <A_n, T_n\_alpha>\}$$

$$\{<A_1, T_1>, <A_2, T_2>, ..., <A_n, T_n>\}$$

If and only if, for all i (i = 1,2, ..., n), type $T_{i\_alpha}$ is the maximal type with respect to type $T_i$, for example:

RELATION {E alpha, R alpha}
/ * Maximal type with respect to the relation type RELATION {E ELLIPSE, R RECTANGLE * /

2. *T/R_omega type*: the type T/R_omega is said minimal type with respect to a *tuple /relation* type T, with headings, respectively:

$$\{<A_1, T_{1\_omega}>, <A_2, T_{2\_omega}>,.., <A_n, T_{n\_omega}>\}$$
$$\{<A_1, T_1>, <A_2, T_2>, ..., <A_n, T_n>\}$$

If and only if, for all i (i = 1, 2, ..., n), type $T_{i\_omega}$ is the minimal type with respect to type $T_i$, for example:

RELATION {E omega, R omega}
/ * Minimal type with respect to the relation type RELATION {E ELLIPSE, R RECTANGLE * /

### III.2. Multiple inheritance

As Date and Darwen it defined, a type lattice exhibits multiple inheritance if and only if that lattice must have at least one type that has at least two immediate supertypes. We address the following examples to show this fact on scalar types, and nonscalar; namly the tuple and relation types.

### III.2.1. Scalar type

An example of *multiple* inheritance is that between SQUARE, RHOMBUS and RECTANGLE types; precisely, SQUARE is a *proper subtype* of the two *immediate supertypes* RHOMBUS and RECTANGLE. In Tutorial D, the type SQUARE can be defined as follows:

```
TYPE  SQUARE
        IS   {RECTANGLE, RHOMBUS
            POSSREP {A = THE _A (RECTANGLE),
                     B = THE _B (RECTANGLE),
                     C = THE _C (RECTANGLE),
                     D = THE _D (RECTANGLE)}}.
   /* THE_  operator allows the user to access the components corresponding to a
                    specified value of type   RECTANGLE */
```

In that definition, it is possible to specify the possrep in term of the type RECTANGLE (as used in the definition above) or the type RHOMBUS, since the type SQUARE is a subtype of the latter two types.

In multiple inheritance models, Date and Darwen have introduced other concepts as:
a. *The least specific type unique:* let T scalar type and v a value of type T, T is the least specific type if and only if the value v is of type T and not of any proper supertype of T.

b. *The most specific type unique*: let value v be of type T, the type T is the most specific type if and only if no proper subtype T of type T exist such that v is also of type T'.

### III.2. Non scalar type

To start with multiple inheritance of nonscalar tuple and relation types, we consider for example the following tuple types:

        TUPLE {E ELLIPSE, R RECTANGLE}     /* type tuple "ER" */
        TUPLE {E CIRCLE, R RECTANGLE}      /* type tuple "CR" */
        TUPLE {E ELLIPSE, R SQUARE}        /* type tuple "ES" */
        TUPLE {E CIRCLE, R SQUARE}         /* type tuple "CS" */

    Types CR and ES are both subtypes of type ER and both supertypes for type CS. Then, then CS is a subtype with two *immediate supertypes*. Thus, any tuple variable of type ER can be with a value that the components E is of type CIRCLE or R is of type SQUARE, or both in same time, this definition is also applied to the relation types.

### IV. Conception of a type system for the orthogonal model of Date and Darwen

It is commonly admitted that the notion of type is essential for databases conception. The description of types is also part of the databases modeling.
   On the one hand, at the language level, the type system provides a conceptual tool to prove the relevance of the structural aspects of language in relation to the typing rules proposed, by classifying expressions according to the types of values witch compute and involve one (or several) type to each value, by examining the relationship between types and expressions.
On the other hand, at the data model level, the type system provides a framework for typing the various concepts of the model and for a uniform definition of its semantics.
In [31] Date and Darwen propose a theory of types to present panoply of simple and complex types namely: scalar types, tuple types and relation types. This proposal has not been structured in a formal system. Specifically, Date and Darwen have not indicated the formalism of the type system that corresponds to the type theory established.
   The objective of our article is to define a type system to represent all types integrated into the type theory proposed by Date and Darwen in [31.32]. This type system constitutes a starting point to determine the formal semantics of the model entirety. For this purpose, we use a simple and rigorous formal structure, by adopting a pseudo-algorithmic approach in order to show the process of inference of the various existing types.

   The formalization of our type system is structured as a standard notation based on a grammatical approach. In addition, our type system is a declarative system accepting two forms of polymorphism: parametric and inclusive. Indeed, we can summarize the characteristics of our type system in the following clauses:

- A declarative approach that takes into account the polymorphism to describe the syntax of the language system.
- A pseudo-algorithmic approach to describe the formal structure of the system.
- A grammatical approach to describe the typing rules.
- The specifications of different types are given under Tutorial D notation.

- The description of the type system is supposed to take into account the inheritance context with these two categories namely: the single and multiple inheritances.
- The support and the consideration of a new particular type for representing the null value.

**IV.1. Typing Principle**

Our aim is to conceive a polymorphic type system with inheritance between types. To do this, we adopt a declarative approach in which the user provides the definitions and type declarations for that the system verifies their consistency.

We believe that the declarative approach, which allows to express the user intention about types in general, is more appropriate for the specification of our type system. This is due precisely, first to the nature of language proposed by Date and Darwen in [31, 32], Tutorial D, that most types are constructed and reported by the user, and secondly, to express types polymorphism induced by a subtyping relationship (as opposed to inferential approaches for which the types are monomorphic and inferred automatically from the program).

The use of a type system imposes restrictions on the form of typing concepts and expressions. Polymorphism [42, 43], i.e the possibility for a term to have multiple data types, provides the finest type systems, imposing fewer restrictions on the user. It is clear that an important property and polymorphic type systems are very popular. For our type system, two kinds of polymorphism are taken into account:

**IV.1.1. Parametric polymorphism**

Parametric polymorphism authorizes the presence of variables or parameters in the types, thereby obtaining an abstract type in relation to these parameters. Alternatively, parametric polymorphism allows the definition of complex types parameterized by types.

In our type system, this form of polymorphism is therefore adapted to the manipulation of abstract data structures such as the user- defined scalar types, tuples and relation types which are defined in terms of one or more components (attributes), each component is a variable declared to be of a scalar or non-scalar. Thus, polymorphism is accomplished in this case when the multiplicity of typing is based on parameters that accept the instantiation of types of parameters. Specifically, in our type system, the concept of parametric polymorphism is performed on scalar types in the fact that the parameters of possible representations Possrep can be of different types for the same scalar type. That is to say, for a given scalar type, types of parameters (components) of each possible representation can be instantiated to any type. This definition by recursion is valid for non-scalar types whose attributes can be of such scalar types. For example the definition of type POINT:

```
TYPE POINT
     POSSREP CARTESIAN {X RATIONAL, Y RATIONAL}
      POSSREP POLAR {R LENTH, THETA ANGLE};
```

Parameters R and THETA are of LENTH and ANGLE types respectively. The type of the parameter R may therefore be instantiated to RATIONAL, LENTH types or others, and the type of the parameter THETA may be instantiated to RATIONAL, ANGLE ... etc. the type POINT is then parametric polymorphic.

**IV.1.2. Inclusive polymorphism**

Inclusive polymorphism is commonly known as subtyping. The definition of inclusion relation between two types allows a term to belong simultaneously to several types. In this kind of polymorphism, we're not talking about instances of types, but inclusion between types in a context of inheritance. For example, consider the declaration of a variable of type ELLIPSE E (in Tutorial D):

VAR E ELLIPSE;

At runtime, the current value of the variable E type ELLIPSE can be any subtype of ELLIPSE, this subtype may be the type CIRCLE. The type CIRCLE is included in the ELLIPSE type, the variable E is of type ELLIPSE and CIRCLE at the same time.

**IV.2. Choice of the formal notation**

The formal structure and logical and mathematical foundation of a model are adequately represented by specified formalisms, most often using the concept of grammars [44]. A grammar G is a sequence of rules that permit defining a syntax and thus a formal language, that is to say a set of words acceptable to a given alphabet. A formal grammar is defined by a quadruplet:

- $\Sigma$ : A finite set of terminal symbols (the alphabet).
- $V$: A finite set of non-terminal symbols.
- S: Start symbol, known as axiom $V \in S$.
- P: A finite set of production rules (pairs consisting of a non-terminal and a sequence of terminals and non-terminal).

For our type system, the specification of every type is given by a grammatical structure; the latter is sensible describing the various elements of our type system. This grammar is defined using standard EBNF notation (*Extended Backus-Naur Form*), we are particularly interested to Wirth definition [45]. The choice of such notation is due to the simplification brought to the description in question, and it seems reasonable to envisage the requirements in terms of the notational matter to provide the adequate formal context in order to represent the type system.
Our type system is sensible represent the whole types of the object / relational model defined by Date and Darwen.

**IV.3. Introduction of type '#'**

In a database model, data logically assimilate what is inspired from reality - name and surname of a person for example, are represented by columns of type character - the idea is to reflect the real world in a logically consistent model. In this sense, the proscription of reject of null values in the model of Date and Darwen seems contradictory compared with reality in which a value may be unknown, even indefinite [46] - the date of marriage of unmarried person for example is not defined, because it is impossible to determine the corresponding field- hence the necessity of a particular type for expressing the null values.
   For this, we will enrich the types of our type system by particular type # [47]; however, we need that type to establish the types semantics in which it will be useful for functions partiality. The fourth proscription in Date and Darwen's model [31, 32] has promoted refutation of any concept of a relation having attributes without values. Thus, support of the type # seems paradoxical with that principle; but its utilization appears also indispensable, based on a logical and semantics approach which inclines the existence of such value in a database [47].

The existence of the type # incites strongly the use of an algebraic grammar [44] where a non terminal term can produce the empty word. Indeed, a grammar G is called algebraic if all its productions rules are of the form:

$$A \rightarrow \beta \text{ avec } A \in V \text{ et } \beta \in (V \cup \Sigma)^*$$

A rule of the form $A \rightarrow \varepsilon$ (empty string) is allowed in the grammar G, such rules are very useful because they allow certain possibilities without forcing.

**IV.4. A Type system for Date and Darwen's Model**

In this section, we define a type system for the model of Date and Darwen, this type system is given as an algebraic grammar $G_{sysT}$. Next, the specification and definition of different types are detailed in a pseudo algorithm. First, this algorithm is meant to describe the types in the context of inheritance with both single and multiple, and outside the context of inheritance in a second.

Indeed, we give the complete structure of the grammar of our general type system, specifying its elements: terminals $\Sigma$, non-terminal V, the axiom S and all typing rules P. Our type system has an empty type which represents null values. In the grammar below a null value is denoted by #, and all different type of # is denoted $T'_{sys}$.

The inheritance model proposed by Date and Darwen is sensible take into account the type union, such type considers the existence of all possible subtypes. Remind that a union type may be dummy when it has no possrep, and a union type (`Union_T'`$_{sys}$) that is not dummy (`Dummy_T'`$_{sys}$) is called regular (`regular_T'`$_{sys}$).

By (`non_Union_T'`$_{sys}$,) we mean a type in an inheritance context which may be *the most specific type* of some values, that type is clearly described in a definition without referring the precision UNION. Thus, the type system associated with Date and Darwen's model can be represented by the following general grammar:

$G_{sysT}$ ($\Sigma$, *V*, S, P):

```
Σ = {#, "[", "]", ORDINAL, UNION, "{", "}", "=",
TYPE, POSSREP, GEN, CONSTRAINT, IS, exp,
<possrep name>, <bool exp>,< user scalar type
name>, INTEGER, RATIONAL, CHARACTER, CHAR,
BOOLEAN, TUPLE, RELATION, alpha, omega}.

V = {< T*sys >,<T'sys >, <Union_Tsys
>,<non_Union_Tsys
>,<regular_Tsys>,<Dummy_Tsys>,<scalar_type>,<nonsc
alar_type>,<user_scalar_type>,<built_in_scalar_t
ype>,<user scalar type def>,<user scalar root
type_def>,<user_scalar_nonroot_type_def>,<user_s
calar_type_name>,<possrep_def>,<possrep_componen
t def>, <possrep constraint def>,<possrep name>
```

```
,<is def>,<possrep or specialization
details>,<Additional constraint def>,<derived
possrep def>,<tuple type>,<relation type>,<tuple
type name>,<relation type name>,<scalar Dummy
type >,<nonscalar Dummy type >, <user scalar
Dummy type>,< user scalar root Dummy type >,<
user scalar nonroot Dummy type>,<user scalar
root Dummy type def>,<user scalar nonroot Dummy
type def>,<nonscalar Dummy type*>,<maximal
type>, <minimal type>, <tuple dummy
type>,<relation
dummy_type>,<heading>,<attribute>,<tuple maximal
type>, <relation maximal type>,< tuple minimal
type>,<relation minimal type>}.
```

**S** = {<T$_{sys}$>}
**P:**
```
<T_sys > ::= <T'_sys > | #
<T'_sys >::= <Union_T'_sys >| <non_Union_T'_sys |<T*_sys>
<Union_ T'_sys >::= <regular_ T'_sys >|< Dummy_ T'_sys >
```

### IV.4.1 Type system with inheritance

We present an alternative structure using pseudo algorithmic specification of single and multiple inheritance from the definitions of each type union.

### A. Single inheritance

In single inheritance, we consider both types Unions, particularly, we dedicate the regular and dummy types, with these two special types alpha and omega.

### A.1. Regular type

A regular type is a union type that has normally one or more possible representations. Note that the specification 'ORDINAL' describes a type with the property that the operator ">" is defined for each pair of values of this type, for example, the type INTEGER is ordinal, by contrast, the type POINT (defined in section II.2.1) wouldn't be an ordinal type, because the notion of one point being somehow greater than another makes no sense.
   Let the immediate supertype of the nonroot type being defined be IST. In that grammar, the <possrep def list> must be specified if IST is a dummy type ; however, it must be empty if and only if the nonroot type being defined is a dummy type. the <Additional constraint def> must be specified if IST is not a dummy type. Also, <Additional constraint def> must be specified ( and must not be empty) unless IST is either a dummy type or a system-defined type without a paossrep.

   Our type system has two type constructors: TUPLE and RELATION, these constructors are instead generators GEN of nonscalar types namely tuple and relation types, respectively.
We declare a variable *entity* of type $T_{sys}$, the inference of a regular scalar type is shown in the pseudo algorithm AlGO$_{Tsys}$, as follows :

/ * Start of the pseudo algorithm AlGO$_{Tsys}$ * /

**Input:** `entity` of `T`$_{sys}$

**BEGIN**

  **IF** ($T_{sys} <> \#$) **THEN**

    **IF** (*Inheritance_context*) **THEN**

      **IF** (*Single_inheritance*) **AND** ( *T'$_{sys}$ is Union*) **THEN**

        **IF** (*T'$_{sys}$ is regular*) **AND** (*T'$_{sys}$ is ordinal*) **THEN**

```
<regular_ T'sys >::= <scalar type>|<nonscalar type>
<scalar type>::=<user scalar type>|<built-in scalar type>
<user scalar type>::= <user scalar type def>
<user scalar type def>::= <user scalar root type
def>|<user scalar nonroot type def>
<user scalar root type def>::= TYPE <user scalar type
name>
[ORDINAL] UNION <possrep def list>
/* The specification UNION indicate a union type, the
specification ordinal is optional*/
<possrep def>::= POSSREP [ <possrep name> ]{ <possrep
component def commalist>[ <possrep constraint def> ] }
<possrep component def>::= <possrep component name>
<regular_T'sys>
<possrep constraint def>::= CONSTRAINT <bool exp>
<user scalar nonroot type def>::= TYPE <user scalar type
name> [ORDINAL] UNION <is def>
/* Definition of a subtype by the specification <is
def> to specify the supertype*/
<Is def>::= IS {<user scalar type name> <possrep or
specialization details>}
<possrep or specialization details>::= <possrep def
list>| <additional constraint def> [ <derived possrep def
list> ]
<Additional constraint def>::= CONSTRAINT <bool exp>
<derived possrep def>::= POSSREP[<possrep name>]{<derived
possrep component def commalist>}
<derived possrep component def>::= <possrep component
name>= <exp>
<Built-in scalar type>::= INTEGER | RATIONAL |CHARACTER |
CHAR | BOOLEAN
<nonscalar type>::= GEN(TUPLE)|GEN(RELATION)
/* Defenition of both constructors TUPLE et RELATION */
GEN(TUPLE)::= <tuple type>
GEN(RELATION)::= <relation type>
<tuple type>::= <tuple type name>
<tuple type name>::= TUPLE <heading>
<relation type>::= <relation type name>
<relation type name>::= RELATION <heading>
<heading> ::= { <attribute commalist> }
<attribute> ::= <attribute name> < regular_ T'sys >
```

        **END IF**

**Output:** *Entity* typée $T_{sys}$ = *root scalar type name | nonroot scalar type name root nonsacalar type name| nonroot nonscalar type name*

## A.2. Dummy type

A dummy type is a union type that has no possrep; in particular, it has no constraints on that possrep. We discuss in this section the dummy scalar and nonscalar types (tuple and relation types).

As mentioned in the definition (7), Date and Darwen have considered both dummy types alpha and omega. However, alpha and omega are not be defined in Tutorial D, at least they are not permitted to be the declared type of anything. The definition of a dummy type is described in ALGO$_{Tsys}$ as follows:

  **ELSE**

    **IF** (*$T_{sys}$ is Dummy*) **THEN**

```
<Dummy_T'sys>::= <scalar Dummy type >|<nonscalar Dummy type >
<scalar Dummy type>::=<user scalar Dummy type>|alpha |omega
/* a scalar dummy type may be without Possrep, orboth special types alpha and omega*/
<user scalar Dummy type>::= < user scalar root Dummy type >|< user scalar nonroot Dummy type>
<user scalar root Dummy type>::= <user scalar root Dummy type def>
<user scalar root Dummy type def>::= TYPE <user scalar type name> UNION
 /* The UNION specification without indicating the possible representation */
<user scalar nonroot Dummy type def>::= TYPE <user scalar type name> UNION  <is def >
<Is def>::= IS {<user scalar type name>}
/* the specification IS without possible representation*/
<nonscalar Dummy type>::=<nonscalar Dummy type*>|<maximal type>| <minimal type>
/* A nonscalaire dummy type can be dummy whose attributes are all of dummyt type, or both maximal and minimal types*/
<nonscalar Dummy type*>::=<tuple dummy type>|<relation dummy type>
<tuple dummy type>::= TUPLE <heading>
<relation dummy type>::= RELATION<heading>
<heading> ::= { <attribute commalist> }
<attribute> ::= <attribute name> <Dummy_T'sys>
```

The dummy types of our type system, alpha and omega for scalar types, and maximal and minimal types for non-scalar types are particulars in the sense that these types are defined in the inheritance model of Date and Darwen by a purely conceptual point of view to describe an overall type hierarchy. We devote the following sections to discuss these different types.

### A.2.1 Type Alpha

We do not think it is possible to define alpha, because we would have to star with:

```
TYPE alpha UNION;
```

Then, any scalar type must be defined as an immediate subtype for alpha; though, built-in scalar types are already being built by the system, then it is not possible to redefine them - for example, specifying the type INTEGER as an immediate subtype for alpha -.

### A.2.2 Type Omega

Theoretically, the definition of omega is possible but not feasible in practice. We would need this:

```
TYPE omega IS {CHAR, INTEGER, RATIONAL, CIRCLE,...};
```

In this case, we should precise all the scalar types names of all the other existing leaf scalar types, and then what if we wanted to define another immediate supertype of omega? We noted that omega can arise as the most specific type of an attribute of an empty relation (Définition4). Then, it is never possible to define a variable to be of type omega, because omega has no values and a variable must always be assigned a value.

Indeed, alpha and omega types are not explicitly defined; they are used by a purely conceptual point of view in order to complete inheritance model of Date and Darwen. However, some other models consider alpha and omega, respectively, as universal and empty types, this alternative has been the subject of the project Muldis Rosetta [48] of Darren Duncan. Note that the omega contains no value; it is qualified to be the empty type.

### A.2.3. Maximal and minimal Tuple/Relation types

At the opposite of a scalar type, where just one maximal type that applies to all possible scalar types, the same is not true for tuple and relation types. Rather, there is one tuple and maximal type for each possible tuple type, and one relation maximal type for each possible relation type. More precisely, if two tuple / relation types $tt_1$, $tt_2$ and $tr_1$, $tr_2$, respectively, have no common supertype, then the corresponding maximal tuple/relation types are distinct, and if they have no common subtype, then the minimal tuple types / relation corresponding types are also distinct.

A maximal tuple / relation type *tt_ alpha* and *tr_alpha* respectively, contains all possible tuples / relationsble of type some subtype or supertype of tuple/relation subtype *tt*, *tr* respectively.
On the other hand, a *minimal* tuple type *tt_ omega* may be empty – for example TUPLE {E omega, R omega} is empty because there is no such tuple- , but not necessarily empty, for example, if *tt* is the type TUPLES { }, then *tt_omega* is equal to *tt*, that is to say; the tuple type tt is itself its own minimal type, and it contains exactly one value: namely 0-tuple (the tuple with an empty set of attributes).

Against, the minimal relation type *tr- omega* is definitely not empty; for example, the type RELATION {E omega, R omega} (see Section III.1.2.) contains one value that is an empty relation of that type, if *tr* is the type RELATION { } then *tr_ omega* is equal to *tr*, that is to say *tr* is its own minimal type, which contains two values: TABLE_DEE and TABLE_DUM. Minimal and maximal tuple / relation types are represented in the part of the grammar G$_{sysT}$:

```
<maximal  type> ::=  <tuple  maximal  type> |<relation
maximal type>
<tuple maximal type > ::= TUPLE <heading>
<relation maximal type > ::= RELATION <heading>
<minimal  type> ::=  <tuple  minimal  type>|<relation
minimal type>
<tuple minimal type > ::= TUPLE <heading>
<relation minimal type > ::= RELATION <heading>
<heading> ::= { [<attribute commalist>] }
<attribute>  ::=  <attribute  name>  <alpha>|<attribute
name> <omega>
```

    **END IF**

**Output:** *entity* typée $T_{sys}$ = Dummy Type name

  **END IF**

### B. Multiple inheritance

For multiple inheritance, a type must have at least two immediate supertypes. Indeed, the specification <scalar type name commalist> must contain at least two scalar types names, and derived possible representations must be empty if and only if the leaf type (nonroot) is a dummy type.

    We do not rewrite the whole structure mentioned in section A, rather we describe the specification of multiple inheritance, and the other specifications of grammar remain the same:

  **ELSE**

  **IF** (M*ultiple_inheritance* ) **THEN**

      ⋮

```
<user scalar nonroot type def>::= TYPE <user scalar type
name> [ORDINAL] UNION <is def>
<is def>::= IS { <scalar type name commalist> <derived
possrep def list> }
/* indicate a list of names of scalar types with a list
of  derived  possible  representation  from  that  of  the
supertype*/
```

      ⋮

    **END IF**

  **END IF**

### IV.4.2 Type system without inheritance

By T*$_{sys}$, we mean a type without inheritance feature. Note that without an inheritance context, Date and Darwen do not consider the dummy type, and they impose that a scalar type must have at least one possible representation.

    **ELSE**

  **IF** ¬ (*Inheritence_context*) **THEN**

    **IF** (*T*$^*_{sys}$ *is ordinal*) **THEN**

```
< T*sys >::= <scalar type>| <nonscalar type>
<Scalar  type>::=  <user  scalar  type>|<built-in  scalar
type>
<user scalar type>::= <user scalar type def>
<user  scalar  type  def>::=  TYPE  <user  scalar  type  name>
ORDINAL <possrep def list>
<possrep  def>::=  POSSREP  [  <possrep  name>  ]{  <possrep
component def commalist>[ <possrep constraint def> ] }
<possrep  component  def>::=  <possrep  component  name>  <
T*sys >
<possrep constraint def>::= CONSTRAINT <bool exp>
<Built-in  scalar  type>::=  INTEGER  |  RATIONAL  |  CHARACTER
| CHAR | BOOLEAN
<nonscalar type>::= GEN(TUPLE)|GEN(RELATION)
GEN (TUPLE)::= <tuple type>
GEN (RELATION)::= <relation type>
<tuple type>::= <tuple type name>
<tuple type name>::= TUPLE <heading>
<relation type>::= <relation type name>
<relation type name>::= RELATION <heading>
<heading> ::= { <attribute commalist> }
<attribute> ::= <attribute name> < T*sys >
```

    **END IF**
 **END IF**

**Output:** *Entity* typée *T*$_{sys}$ = *scalar type name| tuple type name| relation type name*

**ELSE**

```
< Tsys >::= #
 /* The  type # contains no values and it represents null values*/
```

**END IF**
**END ALGO**$_{\textbf{Tsys}}$
                  / * End of the pseudo algorithm ALGOTsys * /

### VI. Conclusion

In [31.32], Date and Darwen have proposed a theory said of type to present a panoply of simple or complex types, the purpose was to complete the relational model by that type theory that can serve as a basis for the conception of a model of inheritance and sub-typing; However, Date and Darwen have not defined the formalism of the adjacent type system in their model.

Indeed, we have presented in this paper a pseudo algorithmic and grammatical description of a type system for Date and Darwen's model. We presented a typing general grammar; each type was detailed in a sub pseudo algorithm by an inference grammar of the type in question.

We take into account the null values, such model which proscribes the use of null values seems paradoxical to the classical relational model principles [33, 34] which constitute the kernel of Date and Darwen's model. To do this, we have introduced a new type noted #, that type represents the set of values expressing one or more occurrences of incomplete information in a database. Similarly, we have showed the specification of single and multiple inheritance in Tutorial D, thus different induced definitions.

The type system presented in this paper constitutes a first step in the perspective of the definition of a semantic study of different concepts of the data model proposed by Date and Darwen.